\documentclass[aps, pre, twocolumn,
               superscriptaddress,
               reprint]{revtex4-2}

\usepackage[utf8]{inputenc}
\usepackage[english]{babel}

\usepackage{amsmath}
\usepackage{amssymb}

\usepackage{graphicx}
\graphicspath{{figures/}{/}}
\usepackage{overpic}

\usepackage{hyperref}
\usepackage{xcolor}

\usepackage{import}

\newcommand{\ud}{\ensuremath{\text{d}}}
\newcommand{\ue}{\ensuremath{\text{e}}}

\newcommand{\xen}{\ensuremath{x_\mathrm{en}}}

\newcommand{\pcut}{\ensuremath{p_\text{cut}}}

\newcommand{\degavg}{\ensuremath{k}}
\newcommand{\suscept}{\ensuremath{\text{S}}}
\newcommand{\infect}{\ensuremath{\text{I}}}
\newcommand{\recov}{\ensuremath{\text{R}}}

\newcommand{\nsus}{\ensuremath{n_\suscept}}
\newcommand{\ninf}{\ensuremath{n_\infect}}
\newcommand{\nrec}{\ensuremath{n_\recov}}
\newcommand{\xinf}{\ensuremath{x_\infect}}
\newcommand{\xsus}{\ensuremath{x_\suscept}}
\newcommand{\xrec}{\ensuremath{x_\recov}}
\newcommand{\nsi}{\ensuremath{n_{\suscept\infect}}}

\newcommand{\tinf}{\ensuremath{t^\mathrm{inf}}}
\newcommand{\Ninf}{\ensuremath{N^\mathrm{inf}}}
\newcommand{\strat}{\ensuremath{\mathcal{S}}}

\newcommand{\graph}{\ensuremath{\mathcal{G}}}
\newcommand{\edges}{\ensuremath{\mathcal{E}}}
\newcommand{\nodes}{\ensuremath{\mathcal{N}}}

\begin{document}
    
    \preprint{APS/123-QED}
    
    \title{Self-adapting infectious dynamics on random networks}
    
    \author{Konstantin Clau\ss}
    \affiliation{%
        Technical University of Munich, Department of Mathematics,
        85748 Garching bei München, Germany
    }%
    
    \author{Christian Kuehn}
    \affiliation{%
        Technical University of Munich, Department of Mathematics,
        85748 Garching bei München, Germany
    }%
    \affiliation{%
        Complexity Science Hub Vienna, 1070 Vienna, Austria
    }
    
    \date{\today}

    \begin{abstract}
        
Self-adaptive dynamics occurs in many physical systems such as
socio-economics, neuroscience, or biophysics.
We formalize a self-adaptive modeling approach, where adaptation takes place within
a set of strategies based on the  history of the state of the system.
This leads to piecewise deterministic Markovian dynamics 
coupled to a non-Markovian adaptive mechanism. 
We apply this framework to basic epidemic models (SIS, SIR) on random networks.
We consider a co-evolutionary dynamical network where 
node-states change through the epidemics and network topology changes
through creation and deletion of edges.
For a simple threshold base application of lockdown measures we
observe large regions in parameter space with oscillatory behavior.
For the SIS epidemic model, we derive analytic expressions for the oscillation period 
from a pairwise closed model. Furthermore, we show that there is a link to self-organized
criticality as the basic reproduction number fluctuates around one. We also study the  dependence of our results on the underlying network structure.
    \end{abstract}
    
    \maketitle

    \section{Introduction}

    Complex adaptive systems \cite{hollandComplexAdaptiveSystems1992} 
    often arise when interacting nodes (or agents)
    adapt their dynamics (or strategies)
    due to external influences.
    This occurs for example in socio-economic systems, such as
    epidemic and information spreading 
    \cite{funkModellingInfluenceHuman2010,pastor-satorrasEpidemicProcessesComplex2015},
    markets \cite{sornetteCriticalMarketCrashes2003},
    evolution of languages \cite{steelsModelingCulturalEvolution2011},
    or evolutionary games with environmental feedback
    \cite{tilmanEvolutionaryGamesEnvironmental2020},
    but also in automation problems
    including machine learning \cite{caoSelfAdaptiveEvolutionaryExtreme2012} or control theory
    \cite{astromTheoryApplicationsAdaptive1983}.
    In standard network models adaptive dynamics is often imposed by
    creation, deletion or rewiring of edges between pairs of nodes
    \cite{grossAdaptiveCoevolutionaryNetworks2008},
    some aspects of which have been extended
    to higher-order networks
    \cite{battistonNetworksPairwiseInteractions2020,
        horstmeyerAdaptiveVoterModel2020,
        schlagerStabilityAnalysisMultiplayer2021}.
    From a more general perspective adaptivity is characterized by a
    set of strategies which are selected either by the system itself or by individual agents in the system,
    each leading to potentially different dynamical rules and future evolution.

    The adaption of strategies may be characterized by a change of parameters.
    This occurs for example in piece-wise deterministic Markov processes (PDMPs)
    \cite{benaimStabilityPlanarRandomly2014,
        hieuDynamicalBehaviorStochastic2015,
        liuAnalysisSIRSEpidemic2017,
        hurthRandomSwitchingBifurcations2020}, where
    some parameter of the system
    evolves according to an additional, independent stochastic process.
    Assuming that the parameter space represents a possible set of strategies,
    this is equivalent to a system where some decision maker (randomly) 
    changes the strategy.
    If this parameter crosses a bifurcation the stability of the system
    changes abruptly, typically leading to an entirely different dynamical evolution.
    On the other hand, if parameter changes are influenced by the history of the system and
    occur with some temporal delay usually
    delay equations \cite{kuangDelayDifferentialEquations2012} are considered.
    Examples are biomedical models of cancer evolution 
    \cite{bakerModellingAnalysisTimelags1998,
        villasanaDelayDifferentialEquation2003},
    population dynamics \cite{gopalsamyStabilityOscillationsDelay1992}
    or
    machine learning \cite{grigoryevaOptimalNonlinearInformation2015}.
    It is reasonable to assume that in real world systems both
    effects occur: delayed adaptation depending on the history of the state and
    piece-wise deterministic evolution coupled to sudden
    strategy changes.
    Still, these approaches have so far not been connected.

    Particularly interesting are systems with policy makers
    who adapt either the rules of the system or their own behavior
    according to some, usually complex, evaluation mechanism.
    For example, restrictions and lockdown laws have been enforced and lifted
    during the Sars-Cov-2 pandemics depending on the recent course of the pandemic
    \cite{schlickeiserReasonableLimiting7Day2021},
    and are regularly changed, e.g.,  due to the emergence of new variants
    or available vaccinations.
    Therefore it is reasonable to ask how basic epidemic models behave in the context
    of adaptive policies.
    This is related to recent models for
    the influence of individual risk perception on the epidemic spreading of Sars-Cov-2
    \cite{dongesInterplayRiskPerception2022}, where direct feedback
    mechanisms are considered.

    Epidemic modeling has a long history \cite{pastor-satorrasEpidemicProcessesComplex2015},
    the most famous examples being the SIS and SIR models
    \cite{kermackContributionMathematicalTheory1927,andersonInfectiousDiseasesHumans1992}.
    In these compartmental models the population is divided into 
    susceptible ($\suscept$) and infected individuals ($\infect$), which after recovery become either susceptible again or recover ($\recov$), for SIS and SIR systems, respectively.
    These models are often considered on contact networks,
    where existing links determine possible infections between individuals, see Ref.~\cite{pastor-satorrasEpidemicProcessesComplex2015} and references therein.
    Many extensions have been studied,
    from additional compartements \cite{brauerCompartmentalModelsEpidemiology2008},
    to adaptive rewiring of edges
    \cite{grossEpidemicDynamicsAdaptive2006,shawFluctuatingEpidemicsAdaptive2008},
    adaptive force of infection
    \cite{capassoMathematicalStructuresEpidemic1993,
        donofrioInformationrelatedChangesContact2009,fenichelAdaptiveHumanBehavior2011}, and
    including vaccination
    \cite{britoExternalitiesCompulsaryVaccinations1991,
        shulginPulseVaccinationStrategy1998, donofrioVaccinatingBehaviourInformation2007,
        zamanStabilityAnalysisOptimal2008}.
    Depending on the specific model and parameters, the system state
    typically either approaches
    a stable disease-free or endemic equilibrium,
    oscillates, or shows more complex dynamics.
    Apparently such adaptive models often become analytically unfeasible,
    even if the high-dimensional network dynamics is reduced to the level of differential equations by appropriate closure techniques. 

    In this paper we propose a framework for history
    dependent strategy adaption with piece-wise deterministic dynamics.
    This is applied to simple models of adaptive epidemics
    for which an analytical treatment is possible.
    In particular we consider a threshold-based activation and deactivation of
    edges within a fixed contact network for the examples of SIS and SIR systems.
    We observe and describe a large parameter region with stable oscillations, as well as bifurcations towards stable disease-free and stable endemic states for the case
    of SIS.
    We derive analytic expressions for the period and observe that
    the network topology significantly influences the agreement by
    comparing different types of random networks.

    The paper is structured as follows.
    In Sec.\ref{sec:dynamics} we propose a model of strategy adaption.
    Section~\ref{sec:EoN} recapitulates SIS and SIR models on networks.
    In Sec.~\ref{sec:epidemics-strategy} we apply the adaptive framework
    to SIS and SIR epidemics on networks.
    A summary and outlook is presented in Sec.~\ref{sec:outlook}.

    \section{Dynamics with strategy adaption}
    \label{sec:dynamics}
    
    There are many ways to include adaptivity into dynamical
    systems, e.g., by rewiring rules \cite{grossEpidemicDynamicsAdaptive2006}
    on the network level or
    by state-dependent parameter changes \cite{donofrioInformationrelatedChangesContact2009} on the
    level of differential equations.
    In contrast, here we extend a system with a strategy
    space $\strat$ such that the dynamics
    depends on the currently active strategy $S_i \in \strat$.
    For example, in a system with infectious dynamics, the strategies $S_i$ could refer
    to different combinations of reduced numbers of average contacts $\degavg_i < \degavg_0$,
    and/or increased measures against transmission, $\tau_i < \tau$.
    Alternatively, such a strategy could impose the emergence
    (or deletion) of additional compartments $X \in \{\mathrm{Q}, \mathrm{E}, \mathrm{\tilde{I}}, \dots\}$ into the model, e.g.,
    enforcing quarantine, separately counting exposed but not infectious agents,
    or introducing new variants of the disease.

    It is natural to assume that the chosen strategy depends on some
    observable function $g : \Gamma \rightarrow \mathbb{R}^n$, which evaluates the
    state $x\in\Gamma$ of the system.
    This is in particular relevant if
    the strategies are chosen by some decision maker(s)
    based on their evaluation metric.
    Note that in real social systems often many different metrics are used
    to evaluate the same state, e.g., 
    incidence, hospitalization rate, and vaccination rate are all possible options 
    to evaluate the state of an epidemic.
    For the sake of simplicity we consider a one-dimensional metric, $n=1$, $g(x)\in\mathbb{R}$,
    in the following.

    In general, the history and the current state are important for
    the evaluation of future strategies.
    In order to include the recent history of the state
    we average the observable $g$ over time
    \cite{donofrioInformationrelatedChangesContact2009},
    \begin{equation}
        J(t) = \int_{0}^\infty \rho(t') g[x(t - t')] \ud t'
        = \int_{-\infty}^t \rho(t - t') g[x(t')] \ud t',
        \label{eq:def-indicator}
    \end{equation}
    with a suitable integration kernel $\rho$, also known as delay kernel,
    satisfying $\int_{t = 0}^\infty \rho(t)\ud t = 1$.
    The function $J$ at the time $t$ acts as a measure on how the
    strategy of the system is adapted due to the history of
    the state. We call $J$ strategy function in the following.
    Additionally let us define the adaption function $A$ which specifies
    which strategy $S\in\strat$ is currently chosen.
    For observable $g$ and strategy function $J$ defined as above,
    the adaption function is defined as a mapping from
    $A : T \times \mathbb{R} \times \strat \rightarrow \strat$,
    where $T$ is the time domain and $A(t, J(t), S) = S'\in\strat$.

    We define an adaptive dynamical system with discrete strategies as follows.
    Let $\Phi$ be a map from
    the strategy space $\strat$ to a flow on $\Gamma$,
    i.e., $\Phi_S := \Phi(S)$ is a mapping
    $\Phi_S : T\times \Gamma \rightarrow \Gamma$ for all $S\in\strat$
    with $\Phi_S(0, x) = x$ and
    $\Phi_S[t_2, \Phi_S(t_1, x)] = \Phi_S(t_1 + t_2, x)$
    for all $t_1, t_2 \in T$ and $x\in\Gamma$.
    An adaptive dynamical system is then defined as the tuple
    $(T, \Gamma, \Phi, \strat, A)$, such that for each $S\in\strat$
    the triple $(T, \Gamma, \Phi(S))$ is a dynamical system and
    where the strategy $S_t$ evolves according to
    the adaption function $A$ as defined above.

    We emphasize that the time evolution of the system
    depends on the current strategy $S_{t}$, and simultaneously the
    current strategy depends on the time evolution of the system.
    This leads to a nontrivial feedback mechanism and possibly complex and
    interesting dynamics.
    In order to introduce such an adaptive mechanism to some dynamical system,
    it is therefore necessary to specify a set of strategies $\strat$ and their
    implications on the internal dynamics,
    some observable $g$ and strategy function $J$, and how
    the system adapts its dynamics to changes in $J$ through the function $A$.

    This framework connects piecewise deterministic
    Markov processes and delay equations:
    If the switching of strategies $A$
    is governed by a Markov process
    the full system is equivalent to PDMPs,
    see e.g. \cite{benaimStabilityPlanarRandomly2014,
        hieuDynamicalBehaviorStochastic2015,
        liuAnalysisSIRSEpidemic2017,
        hurthRandomSwitchingBifurcations2020}.
    On the other hand,
    if the selected strategy depends on some parameter 
    evaluated at a delayed time $\tau$ in the past,
    i.e., the integration kernel is
    $\rho(t) = \delta(t - \tau)$,
    the resulting system can be reduced to delay equations
    \cite{busenbergDelayDifferentialEquations1991}.
    In the following we apply this adaptive framework
    to epidemiological models on networks.

    \begin{figure}[b!]
        \includegraphics[scale=0.9]{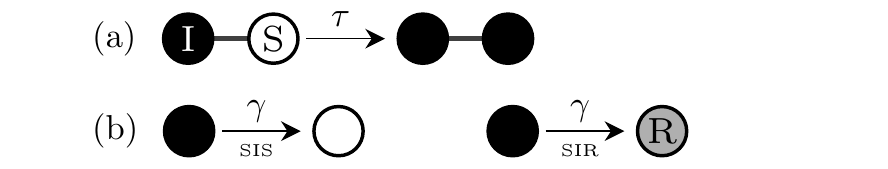}
        \caption{Evolution of SIS and SIR epidemics.
            (a) Infection along $\suscept\infect$-edges with rate $\tau$.
            (b) Recovery of infected nodes with rate $\gamma$
            for SIS (left) and SIR (right).            
        }
        \label{fig:sketch-SIRS}
    \end{figure}
    \section{Epidemics on networks}\label{sec:EoN}

    One of the simplest epidemiological models on networks is the SIS model, see \cite{pastor-satorrasEpidemicProcessesComplex2015}.
    Let $\graph = (\nodes, \edges)$ be a graph consisting of nodes $\nodes$ and
    edges $\edges$.
    The state of each node $n \in \nodes$
    is either susceptible ($X(n) = \suscept$) or infected ($X(n) = \infect$).
    The epidemic is transmitted through the edges $(i,j)\in\edges$
    from infected nodes to adjacent susceptible nodes with a fixed
    transmission rate $\tau$, see Fig.~\ref{fig:sketch-SIRS}.
    Recovery of infected nodes occurs with rate $\gamma$. For SIS
    recovered nodes become susceptible again.
    The full SIS dynamics on a network is described by the unclosed
    differential equations for the number of infected $\ninf$ and
    the number of susceptible $\nsus$ individuals \cite{kissMathematicsEpidemicsNetworks2017},
    \begin{align}
        \begin{split}
            \ninf' &= - \gamma \ninf + \tau \nsi,  \\
            \nsus' &= \phantom{+} \gamma \ninf - \tau \nsi,
        \end{split}\label{eq:sis-unclosed}
    \end{align}
    where $\nsi$ denotes the number of $\suscept\infect$-edges.
    The differential equation for $\nsi$ depends on the higher-order
    moments $n_{abc}$ with $a,b,c \in \{\suscept,\infect\}$, which
    in general leads to an infinite series of differential equations.
    Only for specific network types, such as the complete graph, or by
    applying closure relations to Eq.~\eqref{eq:sis-unclosed}
    one obtains a finite system of differential equations, which in general only approximate
    the full network dynamics, see e.g.\ \cite{kuehnMomentClosureBrief2016,kissMathematicsEpidemicsNetworks2017}.

    In the following we focus on the simplest closure relation, given
    by the pairwise approximation
    $\nsi \approx \frac{\degavg_0}{N} \ninf \nsus$,
    which is based on the assumption that connections
    between $\suscept$ and $\infect$ individuals are spread homogeneously
    through the network.
    Here, $\degavg_0$ denotes the average degree of nodes in the network and
    $N = |\nodes|$ is the number of nodes.
    Without birth and death-processes the number of individuals remains
    constant and is given by
    $N = \ninf + \nsus$.
    Altogether, introducing the proportion of infected
    individuals $\xinf = \ninf / N \in [0, 1]$
    the dynamics reduces to the one-dimensional ODE
    \begin{align}
        \xinf' = - \gamma \xinf + \tau \degavg_0 (1 - \xinf)\xinf.
        \label{eq:ode-pairwise}
    \end{align}
    Stability analysis reveals two different regimes \cite{kissMathematicsEpidemicsNetworks2017}:
    For $\beta := \tau \degavg_0  / \gamma < 1$ the disease free state
    $x_0 = 0$ is stable and it is the only equilibrium within the reasonable interval $\xinf \in [0, 1]$.
    For $\beta > 1$ a transcritical bifurcation occurs and
    $x_0$ becomes unstable, while the endemic state
    $x_e := 1 - 1/\beta \in [0, 1]$ emerges as a stable equilibrium.

    Another simple model for the spreading of a
    single epidemic wave through a population is the
    SIR model,  see e.g. \cite{kissMathematicsEpidemicsNetworks2017}.
    Here, infected individuals which recover
    are fully immune and removed ($\recov$) from the system,
    \begin{align}
        \begin{split}
            \ninf' &= - \gamma \ninf + \tau \nsi,  \\
            \nsus' &= \phantom{+ \gamma \ninf} - \tau \nsi,\\
            \nrec' &= \phantom{+} \gamma \ninf.
        \end{split}\label{eq:sir-unclosed}
    \end{align}
    The corresponding pairwise closed moment equations reduce to the
    two-dimensional system
    \begin{align}
        \xinf' &= -\gamma\xinf + \tau \degavg_0 \xinf \xsus\\
        \xsus' &= \phantom{+\gamma\xinf} -\tau \degavg_0 \xinf \xsus.
    \end{align}
    The  fraction of recovered individuals follows from
    $\xrec = 1 - \xinf -\xsus$ for all times.

    \section{Epidemics with strategy adaption}
    \label{sec:epidemics-strategy}
    We consider an  epidemic system on a network $\graph = (\nodes, \edges)$ with transmission rate $\tau$ and recovery rate $\gamma$.
    The basic network structure of $\graph$ is assumed to be fixed, i.e., the set of edges
    $\edges$ does not change.
    Adaptive activation and deactivation of edges occurs within this framework by
    assigning a weight
    $w :\edges \rightarrow \{0, 1\}$ to each edge $e\in\edges$,
    which determines if the edge is active, $w(e) = 1$, or not $w(e) = 0$.
    More generally, one could assume transmission rates $\tau(e) = w(e) \tau$,
    which are  relevant in the context of different preventive measures, such as quarantine
    \cite{maierEffectiveContainmentExplains2020,kucharskiEffectivenessIsolationTesting2020} or social distancing \cite{giordanoModellingCOVID19Epidemic2020}.
    In order to make this system adaptive we consider the
    simplest case with only two strategies, $\strat = \{S_0, S_-\}$, which determine the
    number of active edges.
    Here, the null-strategy $S_0$ corresponds to case where all (existing) edges in the network
    are active, $w(e) = 1\, \forall e \in \edges$.
    The strategy $S_-$ corresponds to a lockdown strategy, where
    each edge is deactivated with probability $\pcut$, 
    i.e.,  $\mathbb{P}[w(e)\mapsto 0] = \pcut$.
    This means that the average degree in the adaptive network becomes time dependent
    with $\degavg_- = \degavg_0(1 - \pcut)$.

    For the adaptive mechanism $A$ we consider prevalence and incidicence as suitable observable
    functions, i.e.,
    $g_1(t, \xinf) = \xinf$ and
    $g_2(t, \xinf) = \tau \nsi/N \approx
    \tau \degavg(t)\xinf\xsus$.
    Furthermore, the kernel in Eq.~\eqref{eq:def-indicator} is chosen such that $g$ is averaged
    over fixed time intervals
    $[t- \Delta t, t]$ for some time span $\Delta t$, i.e.,
    $\rho(t) = \frac{1}{\Delta t}[\Theta(t) - \Theta(t + \Delta t)]$
    with Heaviside function $\Theta$.
    This leads to
    $J_{1,2}(t) =\int_{t - \Delta t}^t\,g_{1,2}[t', \xinf(t')]\,\ud t'$.
    We emphasize that such an average has been a commonly used measure during
    the Corona pandemic, e.g., in terms of the so-called $7$-day incidence, see e.g.,
    \cite{schlickeiserReasonableLimiting7Day2021}.
    Let us emphasize that $J$ measures the seriousness of the
    current pandemic situation, where more infections or larger
    prevalence manifest in  larger values of $J$.

    The simplest possible adaptive function $A$ for two levels is
    based on fixed thresholds.
    For this, we define two thresholds   $\xi_+ > \xi_-$ and the following strategy adaptions,
    \begin{align}
        \begin{split}
            A(t, J(t), S_0) = \begin{cases}
                S_- & \text{for } J(t) \geq \xi_+, \\
                S_0 & \text{else,}
            \end{cases}\\
            A(t, J(t), S_-) = \begin{cases}
                S_0 & \text{for } J(t) \leq \xi_-, \\
                S_- & \text{else.}
            \end{cases}
        \end{split}
    \end{align}
    This means, whenever the strategy function hits the predefined
    thresholds
    from below or from above, respectively, the system automatically
    enters the state of social distancing or goes back to the usual state.
    Let us emphasize that the memory dependence
    is hidden in the strategy function $J$.
    This function is continuously evaluated
    for all $t$ and thereby determines the current strategy
    $S\in \strat$ of the system.
    Note, that there are numerous possibilities for defining the
    adaption function, which can either deterministically
    or also stochastically select the future strategy.
    In the latter case the system contains two distinct sources of stochasticity,
    one from the dynamics and one from the adaption mechanism, which adds another
    layer of complexity. Here we focus on the deterministic case.

    We emphasize that our proposed model differs quite substantially from recent stochastic SIRS models 
    \cite{hieuDynamicalBehaviorStochastic2015,
        liuAnalysisSIRSEpidemic2017,
        hurthRandomSwitchingBifurcations2020},
    which are piecewise deterministic Markov processes (PDMPs).
    In these systems, the switching of parameters is guided by
    an additional random stochastic process, which is
    independent of the (deterministic or stochastic) dynamics.
    Here, in contrast, the dynamical evolution of the system feeds back into
    the adaptive mechanism.
    If the underlying dynamics is stochastic,
    the adaption of strategies is guided by the (stochastic)
    feedback process implemented with the adaptive function.
    On the other hand, in a purely deterministic setting
    the proposed system can be seen as an extension 
    of the phase-space with the set of strategies,
    $\mathcal{X} \times \strat$, where switching between
    different branches occurs deterministically according
    to the adaptive function $A$.

    \subsection{Results for SIS}
\begin{figure}[t!]
        \begin{overpic}[scale=1.]
            {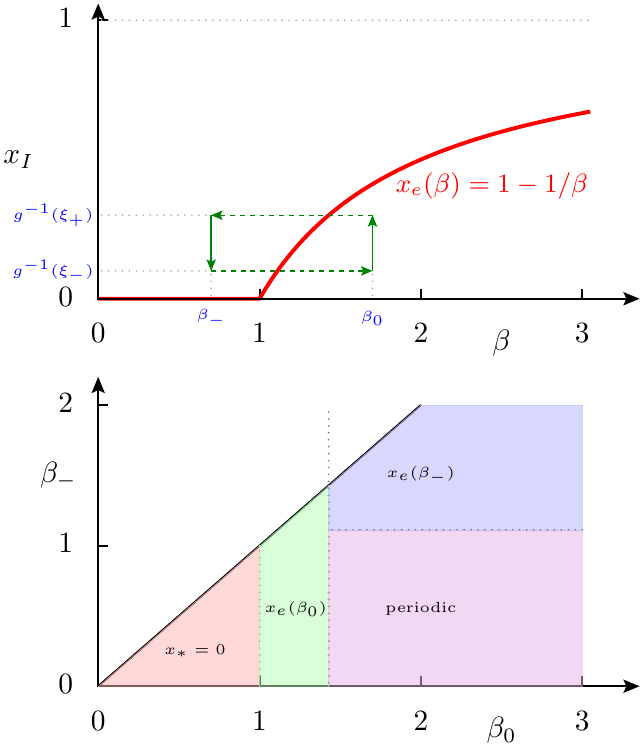}
            \put(-3, 97)  {(a)}
            \put(-3, 47)  {(b)}
        \end{overpic}
        
        \caption{(a) Bifurcation diagram for pairwise closed SIS model with illustration of threshold based periodic dynamics.
            (b) Phase diagram in $(\beta_0, \beta_-)$-plane
            illustrating different regimes of the dynamics for $\Delta t = 0$.}
        \label{fig:skizze-bifurcation}
    \end{figure}

    For the SIS epidemics on a network
    one expects an initial growth of the prevalence $\xinf$
    until $J$ exceeds $\xi_+$. At this point the social distancing
    strategy $S_-$ is applied by removing a proportion of $p_\mathrm{cut}$
    edges from the system, see illustration in
    Fig.~\ref{fig:skizze-bifurcation}(a).
    If sufficienlty many edges are removed, the prevalence decreases
    and $J$ becomes smaller than $\xi_-$, such that the null-strategy
    $S_0$ is reapplied. Within this regime we expect
    periodic behaviour.

    In the following we first derive different dynamical
    regimes in the adaptive pairwise closed SIS model
    and secondly compare this to numerical results
    obtained from simulation on networks.
    Recall that
    for each strategy, $S_0$ and $S_-$, there is one
    stable equilibrium $x_\ast(\beta)$ with
    $x_\ast(\beta) = x_0 = 0$ for $\beta = \degavg \tau /\gamma < 1$ and
    $x_\ast(\beta) = \xen(\beta) = 1 - \beta^{-1}$ for $\beta >1$, see red curve
    in Fig.~\ref{fig:skizze-bifurcation}(a).
    In the following we assume arbitrary but fixed
    thresholds $0 < \xi_- < \xi_+ < 1$ and specify how
    the dynamics depends on the choice of $(\beta_0, \beta_-)$. For an illustration see Fig.~\ref{fig:skizze-bifurcation}(b).
    For simplicity, we also assume that $\Delta t = 0$,
    such that $J(t) = g(\xinf(t))$.
    First, if $\beta_0 < 1$ the state of the system
    converges to the stable equilibrium
    $x_0 = 0$ and the strategy remains in $S_0$ for all times.
    Secondly, for $\beta_0 > 1$ the state of the system
    approaches the endemic state $\xen(\beta_0)$.
    The strategy switches, if $g(\xinf) \geq \xi_+$, which implies the limit $\xen(\beta_0) = 1 - \beta_0^{-1} =  g^{-1}(\xi_+)$.
    In particular,
    for $1 < \beta_0 < \frac{1}{1 - g^{-1}(\xi_+)}$ the
    threshold $\xi_+$ is too large and the state of the system converges to $\xen(\beta_0)$.
    On the other hand, for
    $\beta_0 > \frac{1}{1 - g^{-1}(\xi_+)}$ 
    the strategy will eventually switch to $S_-$ for
    some $t > 0$.
    After switching, the state converges to the stable
    equilibrium $x_\ast(\beta_-)$.
    Similar considerations as above lead to the following
    limits depending on the threshold $\xi_-$. If $\beta_- > \frac{1}{1 - g^{-1}(\xi_-)}$
    the state approaches the stable endemic equilibrium
    $\xen(\beta_-)$, without reaching the lower threshold,
    thus remaining in strategy $S_-$ for all times.
    If $\beta_- < \frac{1}{1 - g^{-1}(\xi_-)}$ the
    stable equilibrium is below the threshold, such that
    after some finite time the system switches back
    to $S_0$ again. Hence, in this regime the dynamics
    is periodic.

    Conversely, if $\beta_0 > 1$ and $\beta_-$ are fixed,
    it is possible to specify limits for the thresholds
    $\xi_\pm$, within which we expect periodic dynamics.
    In particular, one obtains
    $\xi_+ \leq g(1 - 1/\beta_0)$
    and
    $\xi_- \geq g(x^\ast(\beta_-))$.
    With  $\beta_- = \frac{(1 - p_\mathrm{cut}) \degavg_0\tau}{\gamma}$
    these conditions similarly imply a lower bound for
    the cutting probability $p_\mathrm{cut}$ at given $\xi_-$,  given by
    $p_\mathrm{cut} \geq 1 - \frac{\gamma}{[1 - g^{-1}(\xi_-)]\degavg_0\tau}$.

     \begin{figure}[t]
        \includegraphics{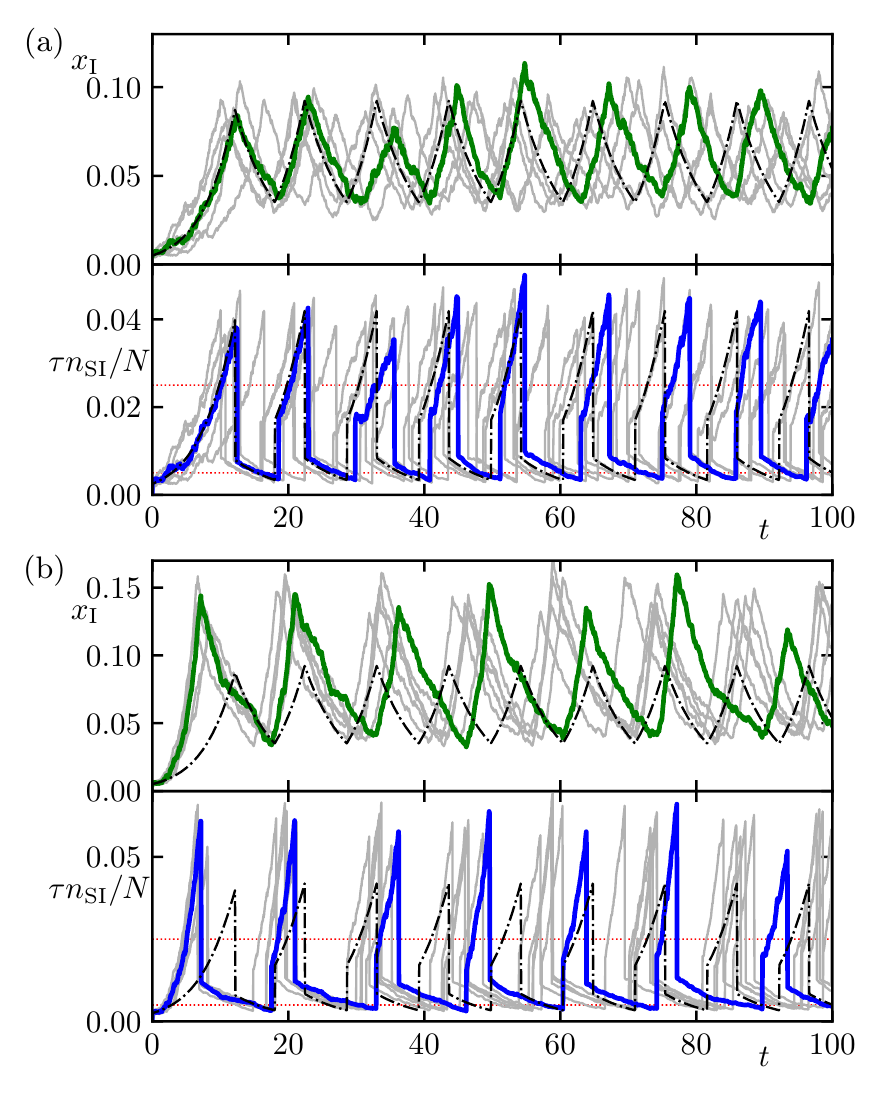}
        
        \caption{
            Relative prevalence $\xinf = I/N$ and incidence
            $\tau \nsi / N$ for SIS epidemics on adaptive network with
            $N=2000$, $\degavg_0=50$, $\gamma = 0.25$, $\beta_0=2$,
            and threshold adaption as in Eq.~\eqref{eq:def-indicator}
            using $g_2$ (incidence thresholds)
            and $\Delta t =5$, $\xi_+=0.025$, $\xi_-=0.005$.
            The lockdown strategy is given by
            $p_\mathrm{cut}(S_-) = 0.8$ ($\beta_- = 0.4$).
            Considered are (a) Erd\"os-R\'enyi networks and
            (b) Barrabasi-Albert networks.}
        \label{fig:sis-continuous-threshold}
    \end{figure}
    In Figure~\ref{fig:sis-continuous-threshold} we illustrate the
    time dependence of the relative prevalence $\xinf$ for the adaptive SIS epidemics
    with strategy function $J_2$
    on two different network types,
    networks form the random Erd\"os-Renyi (ER)  ensemble
    $G(N, p)$ \cite{gilbertRandomGraphs1959,bollobasRandomGraphs2001}
    and scale-free Barabasi-Albert (BA) networks \cite{barabasiEmergenceScalingRandom1999}.
    The parameters are $\gamma = 0.25$, $\degavg_0 = 50$,
    $\beta_0 = 2$ and $\pcut = 0.8$
    with thresholds $\xi_+ = 0.025$ and $\xi_- = 0.005$, i.e., 
    the lockdown strategy is enforced when
    each infected individual infects on average $2.5\%$ of the population
    (over the past time-window $\Delta t = 5$) and it ends below $0.5\%$.
    These parameters correspond to the periodic regime in
    Fig.~\ref{fig:skizze-bifurcation}(b).

    For both network types
    the fraction of infected individuals $\xinf$ oscillates see top panels in Fig.~\ref{fig:sis-continuous-threshold}(a) and (b).
    The corresponding incidence function also oscillates between 
    the predefined thresholds $\xi_\pm$, shown as
    blue curve and red dotted lines, respectively, in the bottom panels.
    Note that the discontinuities of the incidence function are caused by
    the sudden (de)activation of edges, which immediately changes the number of
    $\suscept\infect$-edges and thereby the possible number of new infections.
    For comparison the analytic result of the pairwise closed moment 
    system is shown as a dashed black curve,
    see App.~\ref{app:sis-analytic}.
    This agrees very well with the simulation for the ER-network.
    In contrast, the initial spreading of the epidemics in the BA network 
    is  much steeper, showing also larger maximal values of $\xinf$.
    Consequently, the frequency of fluctuation for these scale-free networks is increased compared to the pairwise closed system.

    \begin{figure}[b]
        \includegraphics{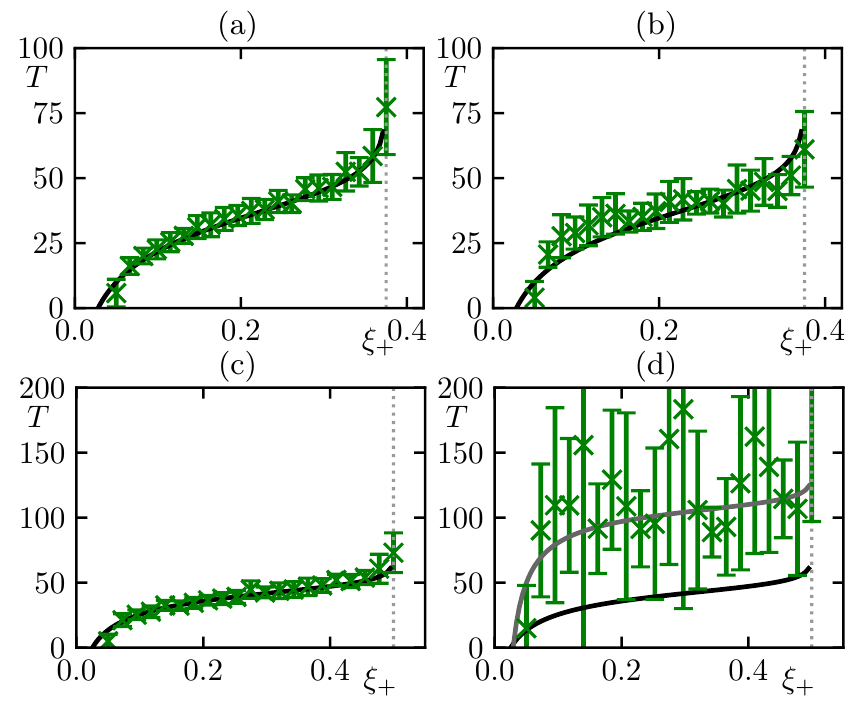}
        
        \caption{
            Period $T$ as a function of $\xi_+$ for
            (a, c) Erd\"os-R\'enyi and
            (b, d) Barrabasi-Albert networks, each with $N = 1000$ nodes and $\degavg_0 = 50$ and
            (a, b) $\beta_0 = 1.6$ and (c, d) $\beta_0 = 2$.
            Other parameters are $\gamma = 0.25$,
            $\xi_- = 0.05$, and $p_\mathrm{cut} = 0.6$.
            Threshold adaption with
            $g_1$ (prevalence thresholds) and $\Delta t = 5$.
            The period is numerically determined from simulation up to
            $t = 1000$ as the average over all time-intervals in which the strategy switches from $S_0$ to $S_-$ and back,
            with $20$ different realizations of the  initial network.
            For comparison the expectation from pairwise closed adaptive SIS model   is shown (black line), in (d) also for $\pcut = 0.49$ (gray line).
            Gray dashed line indicates maximal
            $\xi_+$ for periodic regime,
            see Fig.~\ref{fig:skizze-bifurcation}.
        }
        \label{fig:sis-continuous-period}
    \end{figure}
    
    \paragraph{Period of fluctuations $T$.}
    We further investigate how the period of fluctuations $T$ depends on the parameters of the system for the prevalence strategy function $J_1$.
    The period $T$ is the sum of the expected time $T_{0-}$ to switch
    from $S_0$ to $S_-$ and $T_{-0}$ to switch back.
    For $J_1$ a derivation of $T$ is possible in the pairwise closed
    SIS model, see App.~\ref{app:sis-analytic},
    which leads to
    \begin{align}
        T = T_{0-}& + T_{-0} \nonumber\\
        =\phantom{+}&\frac{1}{\gamma}
        \log \left[\frac{
            \xi_+ (\beta_0 - 1 - \beta_0 \xi_-)}{(\beta_0 - 1) \xi_- - \xi_+ \beta_0 \xi_-}\right]^{\beta_0 - 1}
        \\
        +&\frac{1}{\gamma}
        \log \left[\frac{
            \xi_- (\beta_- - 1 - \beta_- \xi_+)}{(\beta_- - 1) \xi_+ - \xi_- \beta_- \xi_+}\right]^{\beta_- - 1}.
        \nonumber
    \end{align}

    In Figure~\ref{fig:sis-continuous-period} we show $T$ 
    versus the upper threshold $\xi_+$ for two different values
    of $\beta_0 \in \{1.6, 2\}$ and compare network simulations
    (green errorbars) to the analytic result (black curve).
    The upper threshold is chosen in the interval
    $\xi_- \leq \xi_+ < \xen(\beta_0)$.
    Note that the expected period becomes zero at some $\xi_+$ smaller than $\xi_-$,
    see Fig.~\ref{fig:t-vs-xi+}, e.g., for $\Delta t = 0$ this occurs at $\xi_+ = \xi_-$.
    On the other hand, if $\xi_+ \rightarrow \xen(\beta_0)$ the expected period goes to infinity for the analytic model.
    This is also observed in the network simulations:
    For random  ER networks the analytic expression of the pairwise
    closed model fits very well for both $\beta_0$, see (a) and (c).
    For the scale-free BA network and $\beta_0 = 1.6$
    the observed period is systematically too
    large for smaller $\xi_+$ and too small for larger $\xi_+$.
    
    Intuitively, this is explained as follows. The prevalence increases
    initially much faster in scale-free networks due to nodes with very
    large degrees (superspreaders), thereby overshooting the
    threshold significantly stronger, see Fig.~\ref{fig:sis-continuous-threshold}(c).
    Thus, the initial prevalence in the second half of the cycle is increased
    compared to the random networks, thereby increasing the
    period significantly.
    This effect becomes negligible, if the upper threshold gets closer to
    the endemic equilibrium, since here the prevalence typically flattens
    for all network types and the overshooting effect vanishes. Instead, the
    initially much faster speed of infection leads to shorter periods
    in this regime.

    The strong dependence on the topology is revealed when considering
    larger values of $\beta_0 = 2$ and leaving the other parameters unchanged.
    Note that with $\pcut = 0.6$ this still implies $\beta_- = 0.8$ significantly 
    smaller than one.
    In the scale-free network the observed periods are up to four times longer than
    predicted by the moment equations, Fig.~\ref{fig:sis-continuous-period} (d).
    This means that with the same number of deleted edges in scale-free networks
    the progression of the epidemic spreading is slowed down more efficiently
    than in random networks. 
    Note that these longer periods are mainly caused 
    by a much flatter slope  during the lockdown phase $S_-$.
    This corresponds to an effective cutting probability
    of $\pcut \approx 0.49$ (gray line).

    This shows that the network topology leads to observable differences
    in adaptive network systems.
    We believe that such topology induced effects on oscillations can also be observed in more complex models.

    \begin{figure}[b!]
        \includegraphics{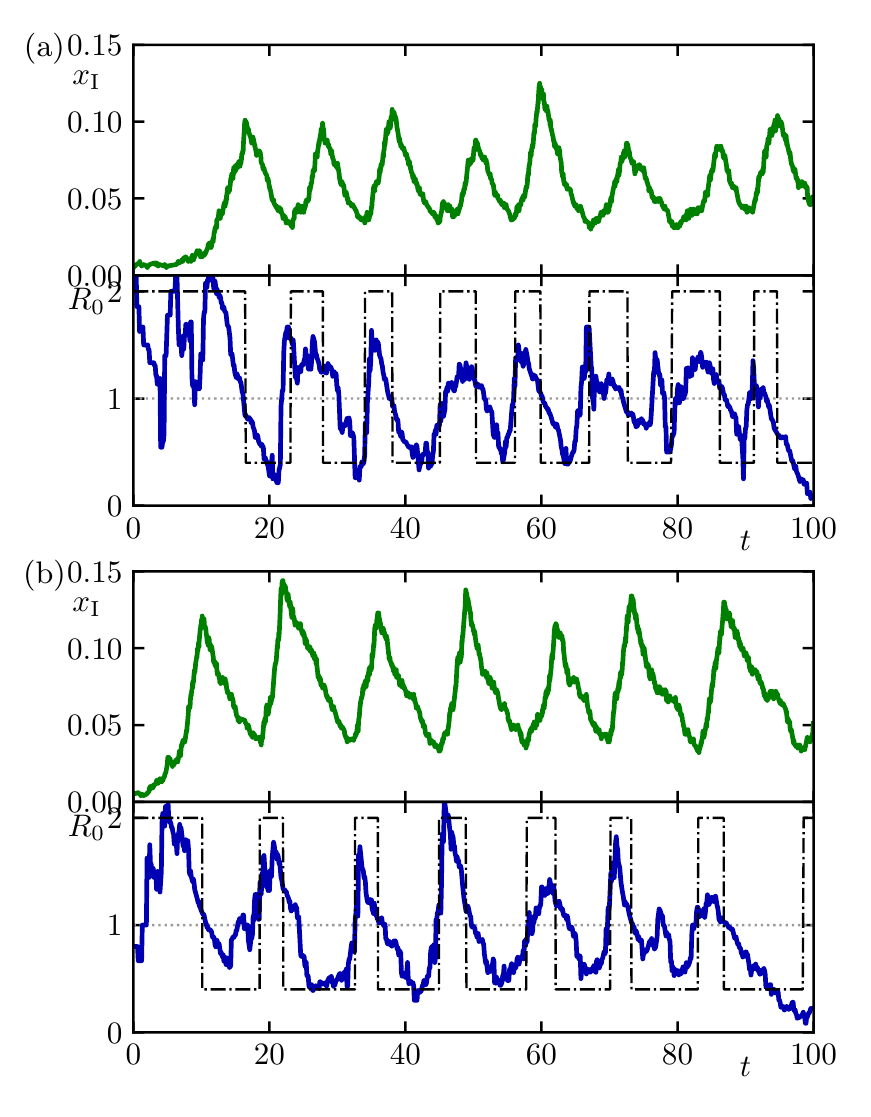}
        \caption{
            Reproductive number $R_0$ as a function of $t$ for
            (a) Erd\"os-R\'enyi and
            (b) Barrabasi-Albert networks, each with $N = 1000$ nodes and $\degavg_0 \approx 50$.
            Other parameters are $\gamma = 0.25$, $\beta_0 = 2$, $\degavg=50$,
            $\xi_+=0.025$,
            $\xi_- = 0.005$, $p_\mathrm{cut}(S_-) = 0.6$, $\Delta t = 5$,
            and the prevalence strategy function $J_1$ is considered.
            The reproductive ratio is numerically determined from the simulation (blue), compared to the prediction of the pairwise model (black dashed).
        }
        \label{fig:sis-reproductive}
    \end{figure}
    \begin{figure}[t]
        \begin{overpic}[scale=0.9, percent]
            {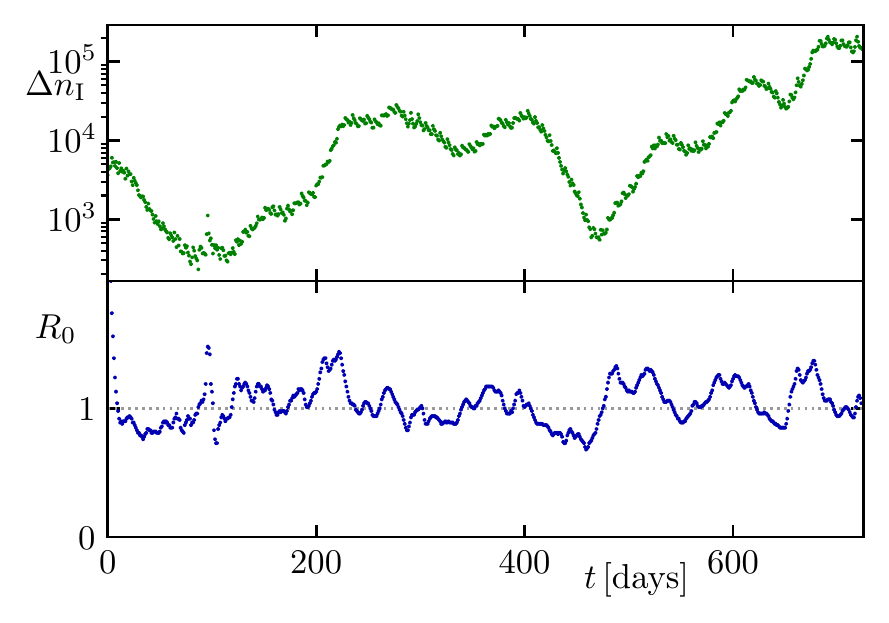}
        \end{overpic}
        \caption{
            Estimation for daily new infections $\Delta \ninf$ and
            $R_0$ based on $7$-day nowcast for Sars-Cov-2 epidemics in Germany between 12.03.2020 and 06.03.2022 \cite{anderheidenSchatzungAktuellenEntwicklung2020,
                anderheidenSARSCoV2NowcastingUndRSchaetzung2022}.
            The nowcast estimation for $R_0$
            is determined by the quotient of new infections for one
            time-interval and those of the preceding interval, assuming
            that each interval corresponds to one generation of the virus.
        }
        \label{fig:reproductive-rki}
    \end{figure}
    \paragraph{Basic reproductive number $R_0$.}
    The basic reproductive number $R_0$ is commonly defined as the expected number of
    new infections caused by one infected individual in a fully
    susceptible community
    \cite{fraserEstimatingIndividualHousehold2007,kissMathematicsEpidemicsNetworks2017}.
    For the pairwise closed SIS model this gives
    the ratio of total rate of infection to total rate of recovery,
    $R_0 = \beta = \degavg \tau / \gamma$ \cite{kissMathematicsEpidemicsNetworks2017}.
    If for infectious disease $R_0 > 1$ an epidemic outbreak
    occurs, while for $R_0 < 1$ the disease vanishes.

    For the SIS epidemics with adaption the average degree on the
    network depends on the currently active strategy.
    Hence, the simplest approximation for $R_0$ is given by
    $\beta \in \{\beta_0, \beta_-\}$ with $\beta_0 = \degavg \tau / \gamma$ and $\beta_- = \beta_0 (1 - \pcut)$.
    Numerically, the reproductive number is obtained by
    averaging the number of infections caused by each infected node over time.
    In particular, let $\tinf_j$ be the time of the $j$-th infection during
    the simulation and let $\Ninf_j$ be the number of infections caused by
    the infected $j$-th node.
    With this, we estimate $R_0$ as
    \begin{equation}
        R_0(t) \approx  \langle \{ \Ninf_j : t - \delta t < \tinf_j \leq t \}\rangle,
    \end{equation}
    where $\langle \cdot \rangle$ denotes the average and we choose the time
    interval according to the recovery time $\delta t = 1/\gamma$.

    In Fig.~\ref{fig:sis-reproductive} we illustrate the
    reproductive ratio $R_0$ as a function of time,
    comparing numerical values (blue) to the simplest estimation based on $\beta$ (black).
    The numerically observed $R_0$ oscillates in a similar pattern around
    $R_0 \approx 1$, however, fluctuating on faster scales and typically 
    below $R_0 = \beta_0 = 2$.
    We assume that the latter is caused by infectious clusters with a small
    number of susceptibles, leading to much fewer infections for large prevalence.

    We emphasize that such fluctuations of the reproductive number
    around $R_0 \approx 1$ are also observed in many different countries during the Corona pandemic,
    e.g., for evolution in Germany see Fig.~\ref{fig:reproductive-rki}.
    This indicates that in real epidemics there is
    some form of self-organized criticality around $R_0 = 1$ 
    due to the conflicting goals
    of reducing the number of infections and increasing the freedom of individuals. 
    In particular, our simple model with adaptive strategies already mimics
    the complex adaptive dynamics due to stricter lockdown measures and more general cautious 
    behavior at higher incidence levels.
    It is therefore reasonable to expect similar observations in more sophisticated models,
    e.g., when each agent chooses from a set of strategies
    and adapts in a more complex manner to the development of an epidemic. Furthermore, our 
    results strongly support the claim that epidemic models with self-limiting feedback 
    mechanisms should be viewed through the lens of self-organized
    criticality~\cite{stollenwerkEvolutionCriticalityEpidemiological2003}.
    
    \begin{figure}[b]
        \includegraphics{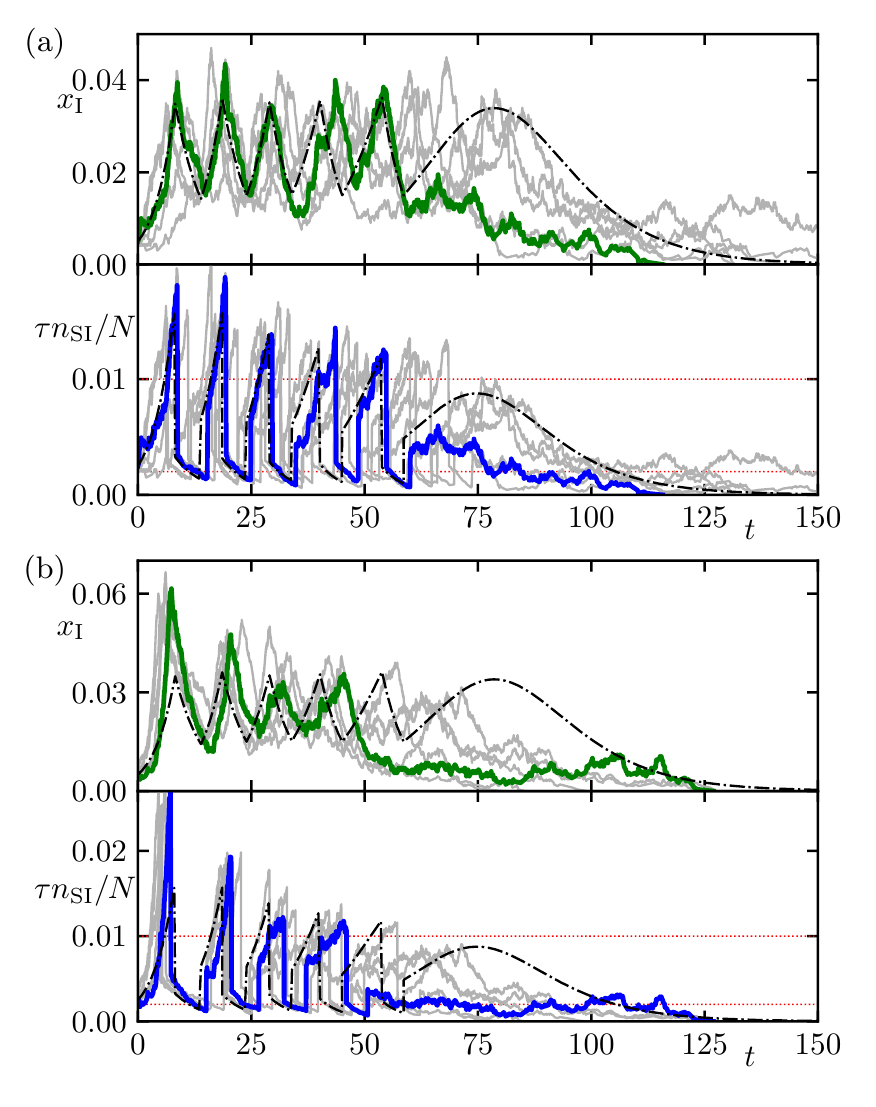}

        \caption{
            Relative prevalence $\xinf = I/N$ and incidence
            $\tau \nsi / N$ for SIR epidemics on adaptive network with
            $N=2000$, $\degavg=50$, $\gamma = 0.25$, $\beta_0=2$,
            and threshold adaption as in Eq.~\eqref{eq:def-indicator}
            using $g_2$ (incidence thresholds)
            and $\Delta t =5$, $\xi_+=0.01$, $\xi_-=0.002$.
            The lockdown strategy is given by
            $p_\mathrm{cut}(S_-) = 0.8$ ($\beta_- = 0.4$).
            Considered are (a) Erd\"os-R\'enyi and
            (b) Barrabasi-Albert networks.}
        \label{fig:sir-continuous-threshold}
    \end{figure}
    \subsection{Results for SIR}
    We further apply the proposed adaptive strategy approach to the SIR epidemics on networks \cite{kissMathematicsEpidemicsNetworks2017}.
    In the SIR model recovered individuals
    do not contribute to the epidemic spreading anymore.
    Similar to the SIS model there is a regime where initial
    transient oscillations are expected.
    When a significant proportion of individuals is
    recovered, $\recov$, the epidemic dies out
    after a certain number of cycles.

    This is illustrated in Fig.~\ref{fig:sir-continuous-threshold}, showing the
    time dependence of relative prevalence and incidence for the SIR
    dynamics on two different network types,
    similar to Fig.~\ref{fig:sis-continuous-threshold}.
    The  considered thresholds are $\xi_+ = 0.01$ and $\xi_- = 0.002$.
    Note that for real epidemics these thresholds should be chosen
    according to the capacity of the health system, which
    typically can sustain only a very small fraction of infected individuals.
    Furthermore, choosing larger thresholds in the simulations
    leads to many infected and recovered individuals already before strategy adaption takes place.
    For both network types we observe the expected initial oscillatory dynamics
    in Fig.~\ref{fig:sir-continuous-threshold}, similar to the SIS system.
    We observe that Erd\"os-R\'enyi networks show comparable
    results to the pairwise closed solution (black dashed line),
    even though individual simulations differ substantially.
    In contrast, the epidemic spreading in the scale-free network takes place
    much faster.
    This is intuitively explained with the power-law
    degree distribution and the existence of a small number of
    highly connected nodes. First, if one of these
    nodes is infected it triggers a lot of subsequent infections.
    Secondly, after its recovery
    it efficiently blocks the epidemic spreading.

    \begin{figure}[b]
        \includegraphics{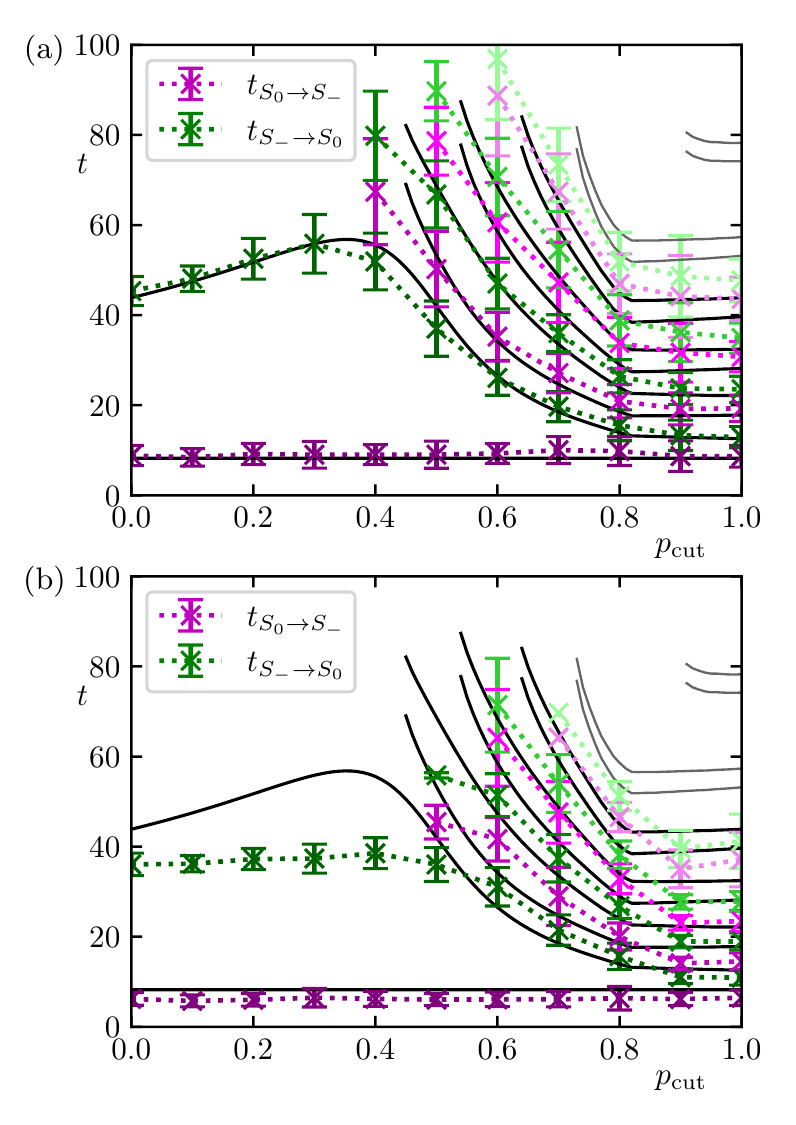}
        \caption{
            Times of switching  as a function of $p_\mathrm{cut}(S_-)$
            for SIR epidemics
            on adaptive network with $N=2000$, $\degavg=50$,
            $\gamma = 0.25$, $\beta_0=2$.
            Threshold adaption as in Eq.~\eqref{eq:def-indicator}
            using $g_2$ (incidence thresholds)
            and $\Delta t =5$, $\xi_+=0.01$, $\xi_-=0.002$.
            Considered are (a) Erd\"os-R\'enyi network
            and (b) Barrabasi-Albert network.
            Colors reporesent
            switch $S_0 \rightarrow S_-$ (violet colors)
            and
            $S_- \rightarrow S_0$ (green colors).
        }
        \label{fig:sir-investigation-pcut}
    \end{figure}

    \paragraph{Adaption times.}
    We investigate numerically how the times of strategy adaption 
    depend on the probability to cut edges during the
    lockdown phase $\pcut(S_-)$.
    This is illustrated in Fig.~\ref{fig:sir-investigation-pcut} for random
    ER and scale-free BA networks
    and compared to numerical solutions 
    of the first order moment SIR system with strategy adaption (black lines).
    For both network types and
    small $\pcut \leq 0.5$ the strategies switch only twice,
    from $S_0$ to $S_-$ (violet colors) and back (green colors).
    In this regime the corresponding $\beta_-  \geq 1$, such that
    the epidemic spreading slows down, but does not revert.
    Note that for $\pcut = 0$ no edges are removed
    and time evolution is equivalent to the
    SIR dynamics on the original networks.
    For increasing $\pcut$ the number of adaptions increases
    and we find very good agreement between simulations on the ER network and the solution of the closed moment equation, see
    Fig.~\ref{fig:sir-investigation-pcut}(a).
    The regime with $\pcut \gtrsim 0.8$ is characterized by initial
    transient oscillations similar to SIS epidemics.
    We observe for the scale-free network that the time
    of the first strategy adaption is much smaller than
    for the closed moment equation.
    This is caused by the initial faster epidemic spreading
    observed in these networks.

    \section{Conclusion and outlook}
    \label{sec:outlook}
    In this paper a self adaptive mechanism with a finite
    set of strategies is proposed, which leads to 
    a coupled system of piecewise deterministic dynamics 
    with history dependent strategy switching.
    Such a modelling approach will be helpful in the description of systems, where the dynamics depends
    strongly on the currently active rules, e.g.,
    social systems and opinion formatting, epidemics.
    It also may be applied to systems, where the full history of the current state changes its evolution,
    like weaker regeneration of skin cells after multiple exposure with sun light leads to increased probability of  cancer.

    This adaptive framework is applied to epidemic models
    on networks.
    For the SIS system we observe a stable regime of
    strategy induced oscillations.
    Their period depends on the network type and
    the degree distribution.
    Based on the pairwise closed model an analytic 
    prediction for the period is derived, which fits very well for Erd\"os-R\'enyi and random regular
    networks and also approximates the period observed in
    scale free networks.

    We emphasize that the proposed adaptive mechanism
    deterministically depends on the state of the system
    and its history. One promising generalization is the to consider a stochastic process for the adaptive mechanism, with state (and history) dependent
    transition rates between the strategies.
    This adaptive mechanism could be coupled with
    deterministic ODE models (like the pairwise
    closed SIS), or with stochastic models (like the
    SIS network model). The latter implies two distinct
    sources of random behavior, which could lead to
    interesting phenomena.
    
    It would be further interesting to investigate in more detail 
    the observation of the fluctuating reproductive rate around its 
    critical value $R_0=1$. There is a clear connection of this observation 
    to self-organized criticality, which
    has been observed in a wide variety of adaptive/co-evolutionary 
    network models~\cite{bornholdtTopologicalEvolutionDynamical2000,kuehnTimescaleNoiseOptimality2012,meiselAdaptiveSelforganizationRealistic2009}.
    We conjecture that having an observable controlling a switching mechanism that entails lowering or increasing
    the infection numbers is the simplest mechanism to obtain self-organized criticality in epidemic
    dynamics. One could now also investigate power law distributions of epidemic event sizes in various models 
    that could further confirm this conjecture. Yet, for our SIS-based model the underlying bifurcation-theoretic 
    mechanism is the switching around a transcritical bifurcation, where one can calculate mathematically that 
    power laws emerge close to the bifurcation point~\cite{kuehnMathematicalFrameworkCritical2011,hurthRandomSwitchingBifurcations2020}. Hence,
    finding evidence for self-organized criticality in very complex epidemic models and/or long-time data
    sets with multiple outbreaks are the most challenging aspects.

    \acknowledgments
    KC and CK thank the VolkswagenStiftung for support via the grant
    ``Self-Dynamics of Self-Adapting Networks'' within a Lichtenberg Professorship awarded to CK. 
    
    \appendix

    \section{Period of threshold adaption for SIS}
    \label{app:sis-analytic}
    In the following we derive an analytic expression for
    the expected period in the oscillatory regime for
    the SIS-model with threshold adaption of strategies.
    For a simplified analysis the pairwise closure,
    of the system is considered, Eq.~\eqref{eq:ode-pairwise}, 
    $\xinf' = - \gamma \xinf + \tau \degavg (1 - \xinf)\xinf$.
    The average node degree takes one of two values,
    $\degavg \in \{\degavg_0,  \degavg_-\}$,
    referring to the strategies $S_0$ and $S_-$, respectively.
    If in $S_-$ the probability to cut links is $p_\mathrm{cut}$, then
    $\degavg_- = \degavg_0 (1 - p_{\mathrm{cut}})$.
    The ratio of the expected rate of infection to the rate of
    recovery, $\beta(\degavg) = \tau \degavg / \gamma$, also takes
    one of two values $\beta \in \{\beta_0, \beta_-\}$ with
    $\beta_- = \beta(\degavg_-) = \beta_0 (1 - p_{\mathrm{cut}})$.

    Solving Eq.~\eqref{eq:ode-pairwise} explicitely for
    initial condition $x(t_0 = 0) = x_0$ one obtains
    \begin{align}
        x(t) = \frac{(\beta - 1)x_0}
        {\beta x_0 + (\beta - 1 - \beta x_0)\ue^{-\gamma(\beta - 1)t} },
        \label{eq:sis-x(t)}
    \end{align}
    where $\beta$ depends on the current strategy and is considered fix between $t_0$ and $t$.
    The threshold criterion for switching from $S_0$ to $S_-$ is
    $J(t_1) = \xi_+$.
    First, we consider $J$ to be the prevalence averaged over
    the past $\Delta t$, which leads to
    \begin{eqnarray}
        J(t_1) &=& \xi_+ = \frac{1}{\Delta t}
        \int_{t_1 - \Delta_t}^{t_1}  x(t') \ud t'\\
        &=&  \frac{1}{\Delta t}
        \int_{t_1 - \Delta_t}^{t_1}
        \frac{(\beta - 1)x_0}{\beta x_0 + (\beta - 1 - \beta x_0)\ue^{-\gamma(\beta - 1)t}},\nonumber
    \end{eqnarray}
    with $\beta = \beta_+$.
    This integral is solved using the indefinite integral
    $\int^t \frac{1}{a + b \ue^{-ct'}}\ud t' = \frac{\log(a\ue^{ct} + b)}{ac}$. With $a = \beta x_0$, $b= \beta - 1 - \beta x_0$ and
    $c = \gamma(\beta - 1)$
    we obtain
    \begin{equation}
        \xi_+\Delta t = \frac{(\beta - 1) x_0}{\beta x_0 \gamma (\beta - 1)}
        \log \frac{\beta x_0 \ue^{c t_1}  + b}{\beta x_0 \ue^{c(t_1 - \Delta t)} + b}
    \end{equation}
    and
    \begin{equation}
        \ue^{c t_1} =
        \frac{(1 - \ue^{\beta \gamma \xi_+ \Delta t})(\beta - 1 - \beta x_0)}
        {\beta x_0 (\ue^{\beta \gamma \xi_+ \Delta t} \cdot \ue ^{-c \Delta t} - 1)}.
    \end{equation}
    An equivalent result is obtained for the reverse switching from
    from $S_-$ to $S_0$.
    Thus, the switching time is given for both cases by the equation
    \begin{eqnarray}
        t_\mathrm{s}&(&x_0, \xi, \gamma, \tau, \degavg,\Delta t)\\ &=&\frac{1}{\gamma(\beta - 1)}
        \log \frac{(\beta x_0 + 1 - \beta)(1 - \ue^{\beta \gamma \xi \Delta t})}{
            \beta x_0 ( 1 - \ue^{\beta \gamma \xi \Delta t} \cdot \ue ^{-\gamma(\beta - 1) \Delta t})},\nonumber
        \label{eq:time-derivation-sis}
    \end{eqnarray}
    where for $S_0 \rightarrow S_-$ one has to plugin
    $\xi_+$, $x_0 < \xi_+$, and $\degavg_0$ and for
    $S_- \rightarrow S_0$ similarly $\xi_-$, $x_0 > \xi_-$
    and $\degavg_-$.
    In both cases we implicitly assumed that
    the switching time is larger than the averaging time $\Delta t$.

    \begin{figure}[t!]
        \includegraphics{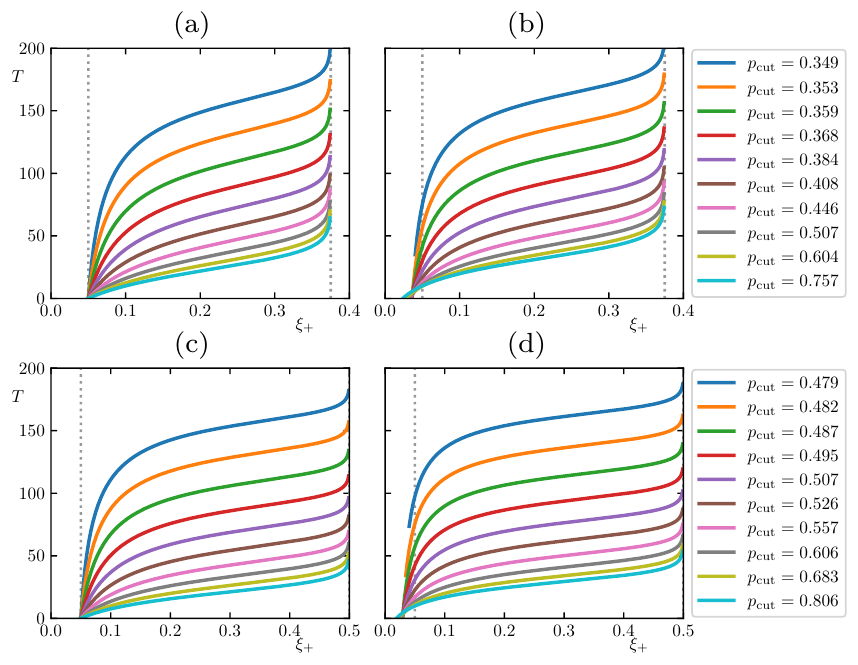}
        
        \caption{Period $T$ as a function of $\xi_+$
            for pairwise closed SIS epidemics with parameters
            $\gamma = 0.25$, $\degavg=50$, $\xi_- = 0.05$
            for different values of $\degavg_- = \degavg (1 - p_\mathrm{cut})$ with
            $p_\mathrm{cut} > p_\mathrm{min} = 1 - \frac{\gamma}{\degavg\tau(1 - \xi_-)}$ as specified.
            Top: $\beta_0 = 1.6$ and (a) $\Delta t = 0$, (b) $\Delta t = 5$.
            Bottom: $\beta_0 = 2$ and (c) $\Delta t = 0$, (d) $\Delta t = 5$. 
            Gray dashed lines indicate $\xi_-$ and maximal $\xi_+$ for  $\Delta t = 0$.
        }
        \label{fig:t-vs-xi+}
    \end{figure}
    In order to derive the full period, we must match
    the initial conditions with the corresponding prevalence values
    $x$ at the last strategy adaption.
    This is evaluated with Eq.~\eqref{eq:sis-x(t)}
    at time $t_{\mathrm{s}}$.
    We obtain
    \begin{equation}
        x(\xi, k, \dots) = \frac{(\beta - 1)(1 - \ue^{\beta \gamma \xi \Delta t})}{
            \beta \ue^{\beta \gamma \xi \Delta t}(\ue ^{-\gamma(\beta - 1) \Delta t} - 1)}.
        \label{eq:xswitch-sis}
    \end{equation}
    Defining $x_+ = x(\xi_+, k_0)$ and $x_- = x(\xi_-, k_-)$ as the prevalences where
    the strategies switch from one to the other, 
    the final result for the period is
    obtained by accordingly plugging $x_\pm$ into Eq.~\eqref{eq:time-derivation-sis} and adding
    both contributions of a full period,
    \begin{equation}
        T = t_s(x_-; \xi_+, k_0) + t_\mathrm{s}(x_+; \xi_-, k_-).
    \end{equation}
    The period depends implicitly on the considered
    network degree $\degavg$ and the cutting
    probability $p_\mathrm{cut}$, the thresholds $\xi_\pm$ and the evaluation time-span $\Delta t$.

    Taking the limit $\Delta t \rightarrow 0$ significantly simplifies these equations.
    For the prevalence strategy function $J_1$ we trivially get the limits
    $x_+ = x(\xi_+, k_0) = \xi_+$ and $x_- = x(\xi_-, k_-) = \xi_-$,
    i.e., the strategy adaption takes place precisely at the threshold levels.
    Equation~\eqref{eq:time-derivation-sis} becomes
    \begin{align}
        t_\mathrm{s}(x_0, \xi) = \frac{1}{\gamma(\beta - 1)}\log \frac{
            \xi (\beta - 1 - \beta x_0)}{(\beta - 1) x_0 - \xi \beta x_0}
    \end{align}
    with $\beta(\gamma, \tau, \degavg)$.
    Altogether, the period is calculated as
    \begin{align}
        T = \frac{1}{\gamma}\Bigg(
        &\log \left[\frac{
            \xi_+ (\beta_0 - 1 - \beta_0 \xi_-)}{(\beta_0 - 1) \xi_- - \xi_+ \beta_0 \xi_-}\right]^{\beta_0 - 1}
        \\ +
        &\log \left[\frac{
            \xi_- (\beta_- - 1 - \beta_- \xi_+)}{(\beta_- - 1) \xi_+ - \xi_- \beta_- \xi_+}\right]^{\beta_- - 1}
        \Bigg)\nonumber
    \end{align}
    Reordering and applying $\beta_- = \beta_0 (1 -  \pcut)$ one gets
    \begin{align}
        T = \frac{1}{\gamma}
        \log \Bigg( \left[
        \frac{\xi_+}{\xi_-}
        \right]^{\pcut\beta_0}
        &\left[
        \frac{\beta_0 - 1 - \beta_0 \xi_-}{\beta_0 - 1 - \beta_0\xi_+}\right]^{\beta_0 - 1}\\
        &\left[
        \frac{\beta_- - 1 - \beta_- \xi_+}{\beta_- - 1 - \beta_-\xi_- }\right]^{\beta_- - 1}
        \Bigg)\nonumber
    \end{align}

    Note that a similar calculation for $J$ being
    the average incidence over past $\Delta t$ leads to
    the condition
    $\xi_+ = \frac{\tau \degavg_0}{\Delta t}
    \int_{t - \Delta_t}^t  x(t')[1 - x(t')] \ud t'$
    which isbecomes difficult to solve analytically.

    In Fig.~\ref{fig:t-vs-xi+} we illustrate the period $T$ as a function of
    $\xi_+$ for different values of $\pcut$ within the periodic regime.
    For smaller $\pcut$ the system converges towards an endemic state, for larger one to
    the disease-free state, see Fig.~\ref{fig:skizze-bifurcation}.
    Increasing (a) $\Delta t = 0$ to (b) $\Delta t = 5$ mainly influences the
    regime with small $\xi_+$.
    For $\Delta t = 0$ lower and upper boundaries of the periodic regime in terms of $\xi_+$ are given
    by $\max\{\xi_-, \xen(\beta_-)\}$ and $\xen(\beta_0)$, respectively,
    indicated by gray dashed lines.

\end{document}